\documentclass[
  aps,
  pra,
  reprint,            
  superscriptaddress, 
  nofootinbib,        
  amsmath,
  amssymb,
  floatfix,            
  nolongbibliography   
]{revtex4-2}

\usepackage{graphicx}  
\usepackage{dcolumn}   
\usepackage{bm}        
\usepackage{ulem}
\usepackage[colorlinks=true, allcolors=blue]{hyperref}
\usepackage[caption=false]{subfig} 
\usepackage{color}     

\begin{document}

\title{Experimental and theoretical studies of hyperfine structures in $^{21}$Na}


\author{Junho Won}
\email{wonder112@ibs.re.kr}
\affiliation{Center for Exotic Nuclear Studies, Institute for Basic Science, Daejeon 34126, Republic of Korea}

\author{Jeongsu Ha}
\email{jeongsu.ha@ibs.re.kr}
\affiliation{Center for Exotic Nuclear Studies, Institute for Basic Science, Daejeon 34126, Republic of Korea}

\author{Deuk Soon Ahn}

\author{Sunghoon Ahn}

\author{Vivek Chavan}

\author{Anastasiia Chekhovska}
\affiliation{Center for Exotic Nuclear Studies, Institute for Basic Science, Daejeon 34126, Republic of Korea}

\author{Gyoungmo Gu}
\affiliation{Center for Exotic Nuclear Studies, Institute for Basic Science, Daejeon 34126, Republic of Korea}
\affiliation{Department of Physics, Sungkyunkwan University, Suwon 16419, Republic of Korea}

\author{Kevin Insik Hahn}
\affiliation{Center for Exotic Nuclear Studies, Institute for Basic Science, Daejeon 34126, Republic of Korea}

\author{Seongjin Heo}
\affiliation{Institute for Rare Isotope Science, Institute for Basic Science, Daejeon 34000, Republic of Korea}

\author{Jangyong Huh}
\affiliation{Center for Exotic Nuclear Studies, Institute for Basic Science, Daejeon 34126, Republic of Korea}

\author{Dahee Kim}
\affiliation{Center for Exotic Nuclear Studies, Institute for Basic Science, Daejeon 34126, Republic of Korea}
\affiliation{Department of Science Education, College of Education, Ewha Womans University, Seoul 03760, Republic of Korea}

\author{Do Gyun Kim}

\author{Dong Geon Kim}
\affiliation{Institute for Rare Isotope Science, Institute for Basic Science, Daejeon 34000, Republic of Korea}

\author{Jung Bog Kim}
\affiliation{Institute for Rare Isotope Science, Institute for Basic Science, Daejeon 34000, Republic of Korea}
\affiliation{Korea National University of Education, Cheongju 28173, Republic of Korea}

\author{Sunji Kim}
\affiliation{Center for Exotic Nuclear Studies, Institute for Basic Science, Daejeon 34126, Republic of Korea}

\author{Yeong Seok Kim}
\affiliation{Institute for Rare Isotope Science, Institute for Basic Science, Daejeon 34000, Republic of Korea}

\author{Yung Hee Kim}
\author{Zeren Korkulu}
\affiliation{Center for Exotic Nuclear Studies, Institute for Basic Science, Daejeon 34126, Republic of Korea}

\author{Donghyeon Kwak}
\affiliation{Institute for Rare Isotope Science, Institute for Basic Science, Daejeon 34000, Republic of Korea}

\author{Jens Lassen}
\affiliation{TRIUMF, Vancouver, BC V6T 2A3, Canada}

\author{Jin Ho Lee}
\affiliation{Institute for Rare Isotope Science, Institute for Basic Science, Daejeon 34000, Republic of Korea}

\author{Jung Woo Lee}
\affiliation{Center for Exotic Nuclear Studies, Institute for Basic Science, Daejeon 34126, Republic of Korea}

\author{Chaeyeong Lim}
\affiliation{Institute for Rare Isotope Science, Institute for Basic Science, Daejeon 34000, Republic of Korea}
\affiliation{Department of Accelerator Science, Korea University, Sejong 30019, Republic of Korea}

\author{Joochun Park}
\affiliation{Center for Exotic Nuclear Studies, Institute for Basic Science, Daejeon 34126, Republic of Korea}
\affiliation{Department of Physics, Hope College, Holland, Michigan 49423, USA}

\author{Ben Ohayon}
\affiliation{Physics Department, Technion—Israel Institute of Technology, Haifa 3200003, Israel}
\affiliation{The Helen Diller Quantum Center, Technion—Israel Institute of Technology, Haifa 3200003, Israel}

\author{Sung Jong Park}
\affiliation{Institute for Rare Isotope Science, Institute for Basic Science, Daejeon 34000, Republic of Korea}

\author{Xesus Pereira-Lopez}
\affiliation{Center for Exotic Nuclear Studies, Institute for Basic Science, Daejeon 34126, Republic of Korea}

\author{Peter Plattner}
\affiliation{CERN, CH-1211 Geneva 23, Switzerland}

\author{Sung Jae Pyeun}
\affiliation{Institute for Rare Isotope Science, Institute for Basic Science, Daejeon 34000, Republic of Korea}

\author{Liss V. Rodr\'{i}guez}
\affiliation{CERN, CH-1211 Geneva 23, Switzerland}
\affiliation{Max-Planck-Institut f\"{u}r Kernphysik, 69117 Heidelberg, Germany}

\author{B. K. Sahoo} 
\affiliation{Atomic, Molecular and Optical Physics Division, Physical Research Laboratory, Navrangpura, Ahmedabad 380009, India}

\author{Taeksu Shin}
\author{Changwook Son}
\affiliation{Institute for Rare Isotope Science, Institute for Basic Science, Daejeon 34000, Republic of Korea}

\author{Yonghyun Son}
\affiliation{Center for Exotic Nuclear Studies, Institute for Basic Science, Daejeon 34126, Republic of Korea}
\affiliation{Department of Physics and Astronomy, Seoul National University, Seoul 08826, Republic of Korea}

\author{Jaehyeon Song}
\affiliation{Institute for Rare Isotope Science, Institute for Basic Science, Daejeon 34000, Republic of Korea}

\author{Laszlo Stuhl}
\affiliation{Center for Exotic Nuclear Studies, Institute for Basic Science, Daejeon 34126, Republic of Korea}

\author{Kyoungho Tshoo}
\affiliation{Institute for Rare Isotope Science, Institute for Basic Science, Daejeon 34000, Republic of Korea}

\author{Hiroshi Watanabe}
\affiliation{Center for Exotic Nuclear Studies, Institute for Basic Science, Daejeon 34126, Republic of Korea}

\author{Yeong Heum Yeon}
\author{Hee Jung Yim}
\affiliation{Institute for Rare Isotope Science, Institute for Basic Science, Daejeon 34000, Republic of Korea}

\author{Hoon Yu}
\affiliation{Republic of Korea Air Force Academy, Cheongju 28187, Republic of Korea}

\date{\today}

\begin{abstract}
We measured the hyperfine structure constants, $A(3s^2S_{1/2})$ and $A(3p^2P_{1/2})$, of the neutron-deficient isotope $^{21}\text{Na}$ using CLaSsy, a setup dedicated to collinear laser spectroscopy at RAON.
The hyperfine structure constants of $^{21}\text{Na}$ were measured to be $103.6(10)_{\mathrm{stat}}(9)_{\mathrm{syst}}$ MHz for $A(3p^2P_{1/2})$ and $954.9(11)_{\mathrm{stat}}(25)_{\mathrm{syst}}$ MHz for $A(3s^2S_{1/2})$. 
A systematic comparison with the state-of-the-art ab-initio relativistic coupled cluster calculations shows the role of higher-order correlation effects such as triple excitations in $^{21}$Na. 
Furthermore, the measurement demonstrates a capability of the CLaSsy setup to conduct collinear laser spectroscopy experiments with a radioactive beam. 
\end{abstract}

\maketitle

\section{Introduction}
\label{sec-1}
\begin{figure*}[!ht]
    \centering
    \includegraphics[width=0.99\linewidth]{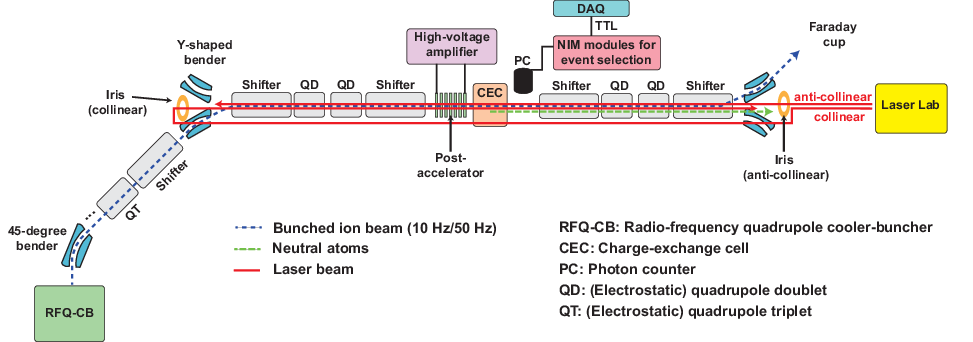}
    \caption{Schematic figure of the CLaSsy setup at RAON. The ion beam is bunched and extracted from the radio-frequency quadrupole cooler-buncher (RFQ-CB) with an energy of 20 keV. The beam passes the two electrostatic benders and enters the CLaSsy beamline. The beam energy is finely tuned at the post-accelerator. The charge-exchange cell neutralizes the ions, and the fluorescence signals are detected by the photon counter. The neutral atoms reaches the end of the horizontal beamline, whereas the remaining positive ions are transported to the Faraday Cup behind the second Y-shaped bender. The periodic time windows synchronized to the RFQ-CB was applied to gain sensitivity for the fluorescence signals. The TTL-like signals are delivered to the data-acquisition (DAQ) system.}
    \label{fig:CLaSsy_setup}
\end{figure*}

The hyperfine structure of atoms originates from interactions between the nucleus and electrons. As the nucleus and electrons are inextricably intertwined, a systematic investigation of hyperfine structure reveals nuclear and atomic properties. 

Since hyperfine interactions originate from the nucleus, their theoretical investigation serves as a robust test of advanced many-body methods used to accurately describe atomic wave functions within the nuclear region~\cite{Sahoo2003,Bijaya4}. These studies also enhance our understanding of the interplay between electron correlation and higher-order relativistic corrections stemming from quantum electrodynamics (QED) effects, achieved through detailed comparisons between experimental measurements and theoretical predictions~\cite{Voltka,Sahoo2018}. Na ($Z=11$) isotopes are of particular interest as atomic and nuclear theory show challenges for the fundamental properties, like nuclear charge radii~\cite{ohayon2022}.

The hyperfine structure constants of $^{23}$Na and $^{22}$Na have been measured with sub-MHz precision using Doppler-free saturated absorption spectroscopy in vapor cells~\cite{pescht1977isotope}. In the case of $^{21}$Na, on-line measurements have been adopted due to its sub-minute half-life ($T_{1/2}=22.4548(53)$ s~\cite{nndc2026}), and the reported precision is lower than that of $^{23}$Na and $^{22}$Na.

In this paper, we present the new measurement of hyperfine structure constants of $^{21}\text{Na}$, employing the collinear laser spectroscopy (CLS) technique~\cite{campbell2016laser,yang2023laser,koszorus2024nuclear}. The experiment was performed at RAON, a new experimental facility dedicated to rare-isotope studies in Korea~\cite{choi2025, chung2023}. The collinear laser spectroscopy setup CLaSsy~\cite{park2025development,lim2026} with a radio-frequency quadrupole cooler-buncher (RFQ-CB) was utilized for the measurement. The hyperfine structure constants $A(3p\,^2P_{1/2})$ and $A(3s\,^2S_{1/2})$ of the stable isotope $^{23}\text{Na}$ were measured first. The measurement served as a validation of the system and, at the same time, allowed us to examine the systematic uncertainties arising from the device performance and the modeling of the fitting function. The same investigation was applied to the case of $^{21}\text{Na}$ to extract the hyperfine structure constants. The measured values were consistent with those in the literature~\cite{1965-Ames,huber1978,touchard1982}. 
By calculating the measured values ab. initio., we emphasize the critical role of electron correlation effects arising from triple excitations—despite their computational complexity—as they are indispensable for attaining theoretical predictions that closely match experimental observations. This is achieved through systematic calculations employing the relativistic coupled-cluster (RCC) method. Additionally, we underscore the importance of incorporating QED corrections to further refine and enhance the accuracy of the theoretical results. 

\section{Experiment}

\begin{figure*}[!ht]\label{fig:23}
    \centering
    \subfloat[Collinear geometry]{
        \includegraphics[width=0.48\textwidth]{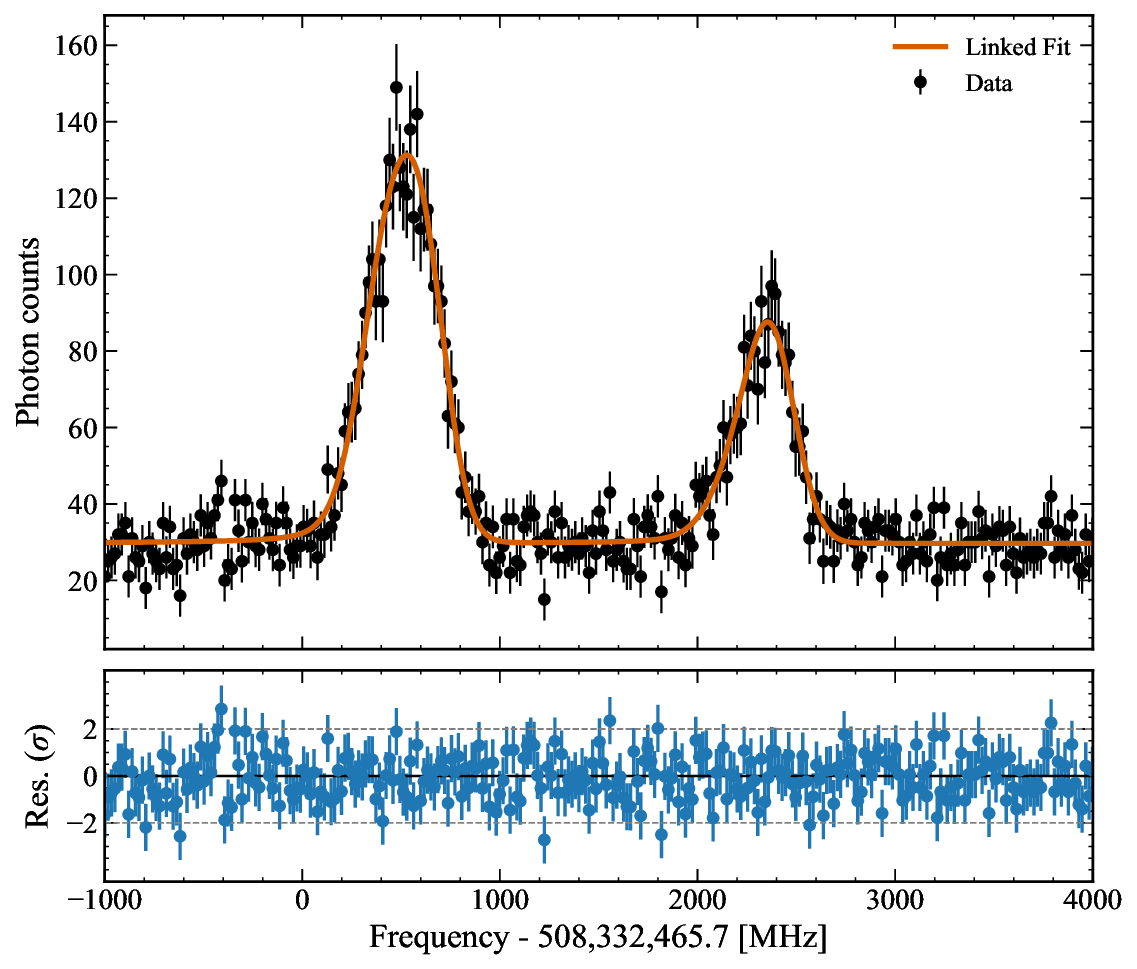}
        \label{fig:23na_co}
    }
    \hfill
    \subfloat[Anticollinear geometry]{
        \includegraphics[width=0.48\textwidth]{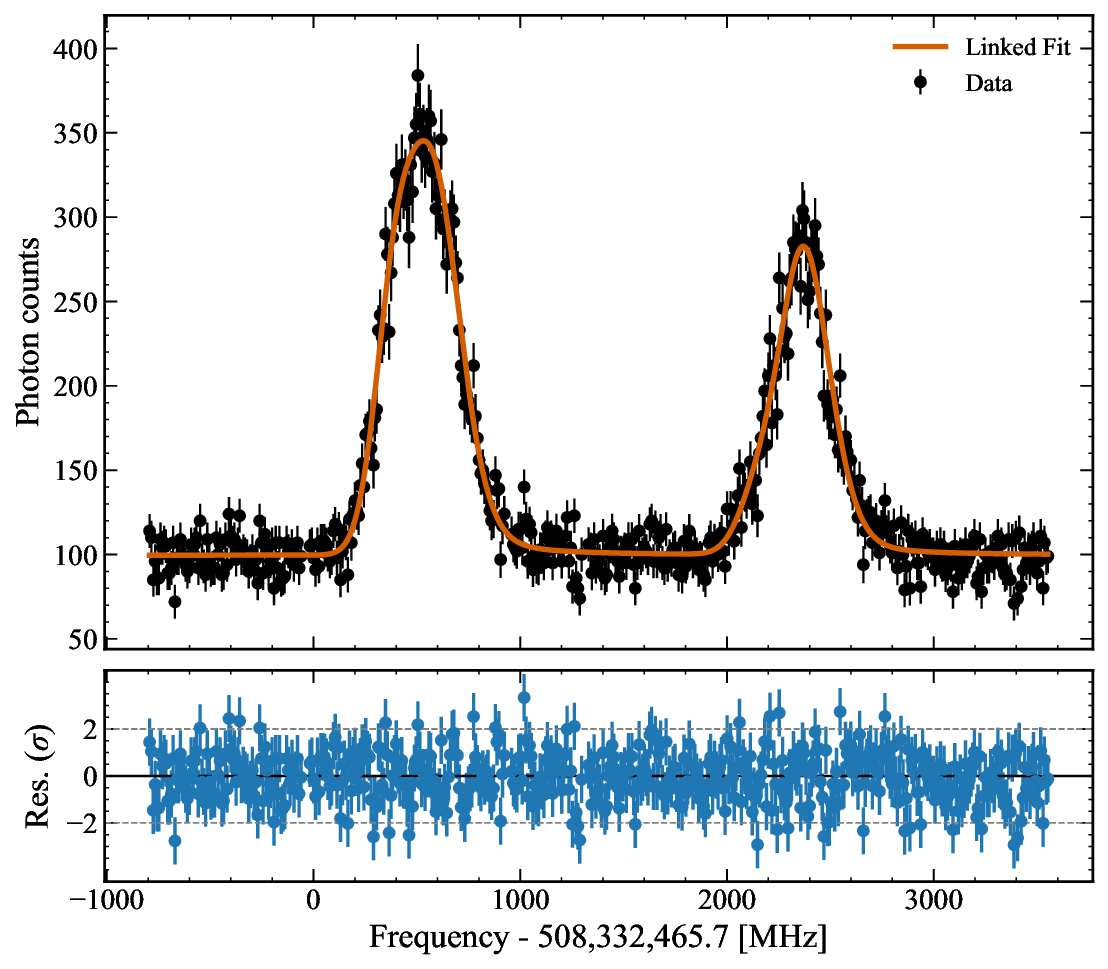}
        \label{fig:23na_anti}
    }

    \caption{\label{fig:23Na_hfs} Hyperfine spectra of the $^{23}\text{Na}$ D1 transition ($3s\,^2S_{1/2}\leftrightarrow 3p\,^2P_{1/2}$) measured with a bunched ion beam from the radio-frequency quadrupole cooler--buncher. The panels show two scans selected from the full dataset of 10 scans: one for (a) collinear geometry, and the other one for (b) anti-collinear geometry. A  frequency offset (508,332,465.7 MHz) is adopted for convenience. Orange solid lines indicate skewed Voigt-profile fits to the hyperfine components. The peak heights were treated as free parameters in the fitting. The bottom panels display the residual plots normalized to the number of counts. The error bars show $\pm 1\sigma$ uncertainty of the counts.}

\end{figure*}

The measurements were carried out at RAON, the new heavy-ion accelerator facility in Korea. The CLaSsy beamline~\cite{park2025development} dedicated to collinear laser spectroscopy was used for the measurements.  A schematic layout of the experimental setup is shown in Fig.~\ref{fig:CLaSsy_setup}. Beams of stable $^{23}$Na and unstable $^{21}$Na were produced using the isotope separation on-line (ISOL) technique. A silicon carbide (SiC) target was irradiated with a $12~\mu\text{A}$, $70~\text{MeV}$ proton beam delivered from the RAON cyclotron~\cite{yeon2025review}. Reaction products were extracted as singly charged ions from surface ionization ($\text{Na}^+$), mass-separated by a dipole magnet, and transported to the spectroscopy beamline~\cite{hashimoto2024design}.

The continuous ion beam was accumulated, cooled, and bunched in the RFQ-CB before being delivered to the CLaSsy beamline. The ions were extracted from the RFQ-CB at a nominal kinetic energy of $20~\mathrm{keV}$ for both $^{21}$Na and $^{23}$Na. The RFQ-CB was operated at 10~Hz for the radioactive $^{21}$Na measurements, and the $^{23}$Na reference spectra were recorded at 10~Hz and 50~Hz depending on the available beam intensity and background conditions. 

Resonance spectra of the Na D1 transition were recorded using the voltage scan method~\cite{kaufman1976}. A scanning voltage was applied to the post-accelerator upstream of the charge-exchange cell (CEC) to tune the ion velocity and shift the resonance condition in the laboratory frame. The voltage was amplified by a high-voltage amplifier, and the output voltage was divided by a high-voltage divider for an accurate read-back of the applied potential \cite{lim2026}. 

Downstream of the post-acceleration stage, the $\mathrm{Na}^+$ bunches entered the heated CEC containing rubidium vapor kept at $250~^\circ\mathrm{C}$. In the CEC, the ions were neutralized with an efficiency of approximately $50\%$~\cite{park2025development}. Details of the CEC design and operating parameters, including rubidium vapor handling and cell geometry, are described in Ref.~\cite{park2025development}. The resulting neutral atom bunches were transported to the interaction region and overlapped with the laser beam. The measurements were carried out in collinear and anticollinear geometries, with the laser propagation direction reversed along the same beamline axis. 

Resonant excitation of the Na D1 transition ($3s\,^2S_{1/2} \leftrightarrow 3p\,^2P_{1/2}$) was provided by a tunable single-mode continuous-wave dye laser (Matisse DX), which was linearly polarized and operated at $589~\mathrm{nm}$ with a short-term spectral linewidth of approximately $100~\mathrm{kHz}$. The laser frequency was monitored by HighFinesse WS-U wavemeter referenced to a stabilized He--Ne laser, and locked via the feedback error signal of the stabilization loop. The frequency was recorded at every data point throughout all scans, and the observed long-term drift was $1.6~\mathrm{MHz/h}$ during data acquisition. 
The optical power was adjusted using a half-wave plate and polarizing beamsplitter.

Although the natural linewidth of the Na D1 transition is about 9.76~MHz~\cite{steck2003}, the estimated laser intensity in the interaction region leads to non-negligible power broadening \cite{Foot}. For the operating conditions, an optical power of 310--350~$\mu$W and a beam diameter of about 0.5--1.0~mm, the standard relation
\begin{eqnarray}
\Gamma_{\rm eff}=\Gamma_0\sqrt{1+I/I_{\rm sat}},
\end{eqnarray}
where $\Gamma_{\rm eff}$ is the power-broadened Lorentzian full-width at half maximum (FWHM$_{L}$ hereafter), $\Gamma_0$ is the natural linewidth, 
$I$ is the laser intensity, and $I_{\rm sat}$ is the saturation intensity, gives an estimated Lorentzian FWHM of approximately 26.4--53.1~MHz. This range was therefore used in the line-shape systematic study described in Sec.~III C.

Fluorescence photons emitted from the excited atoms were collected by a 2-inch-diameter Fresnel lens and imaged onto a photon counter~\cite{lim2026}. The photon counter provided TTL-type pulse outputs, which were counted by the data acquisition system within a time gate synchronized to the RFQ-CB bunch timing to suppress background from scattered laser light and detector dark counts. For each $\sim 30~\mu\text{s}$ bunch, the counting gate was opened for a $20~\mu\text{s}$ window centered on the expected arrival time of the fluorescence signal. 

Photon counts were accumulated for $1\,\text{s}$ per voltage step for the $^{23}\text{Na}$ measurements, whereas the accumulation time was increased up to $4~\text{s}$ per step for $^{21}\text{Na}$ to compensate for the lower radioactive-beam intensity.

\section{Results and Discussion}
\subsection{Hyperfine structure of $^{23}\text{Na}$} 

\begin{figure*}[ht]
    \centering
    \subfloat[Collinear geometry]{
        \includegraphics[width=0.48\textwidth]{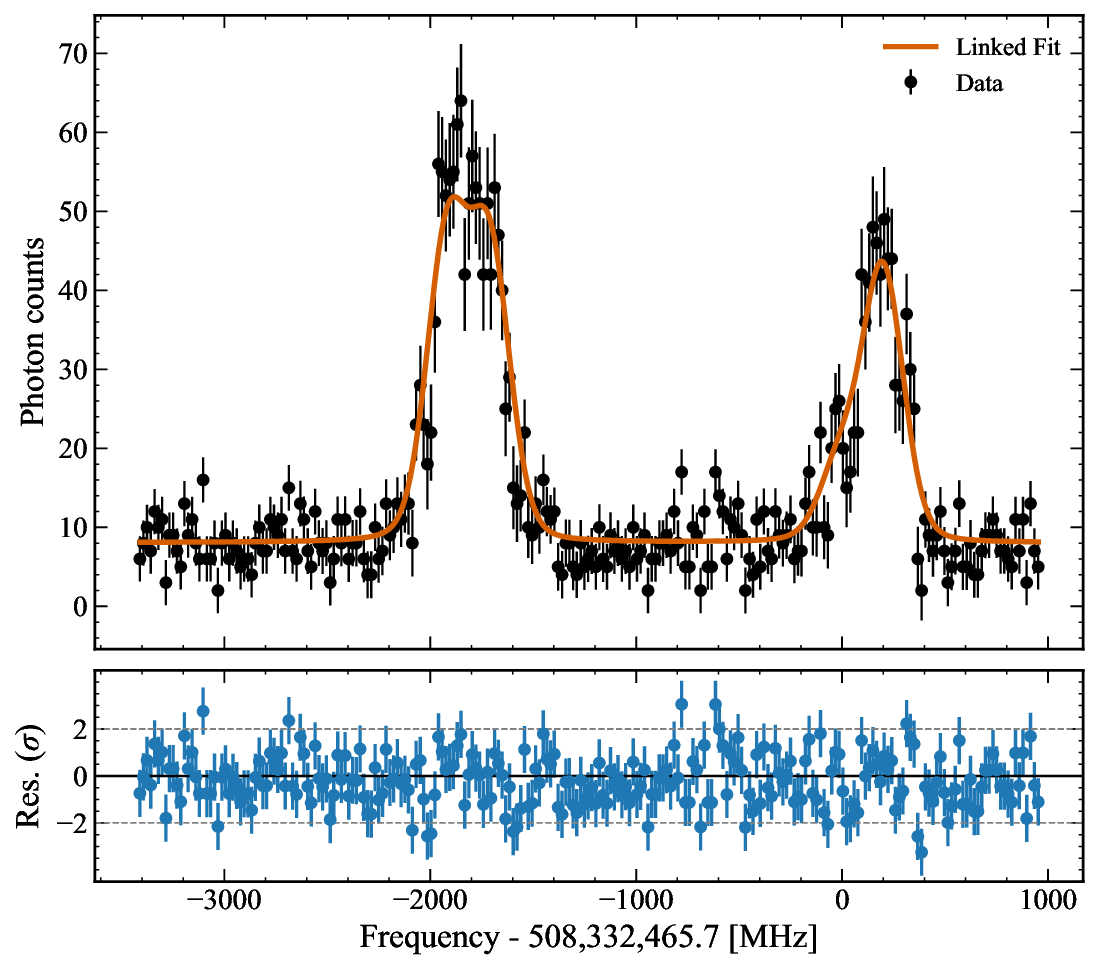}
        \label{fig:21na_co}
    }
    \hfill
    \subfloat[Anticollinear geometry]{
        \includegraphics[width=0.48\textwidth]{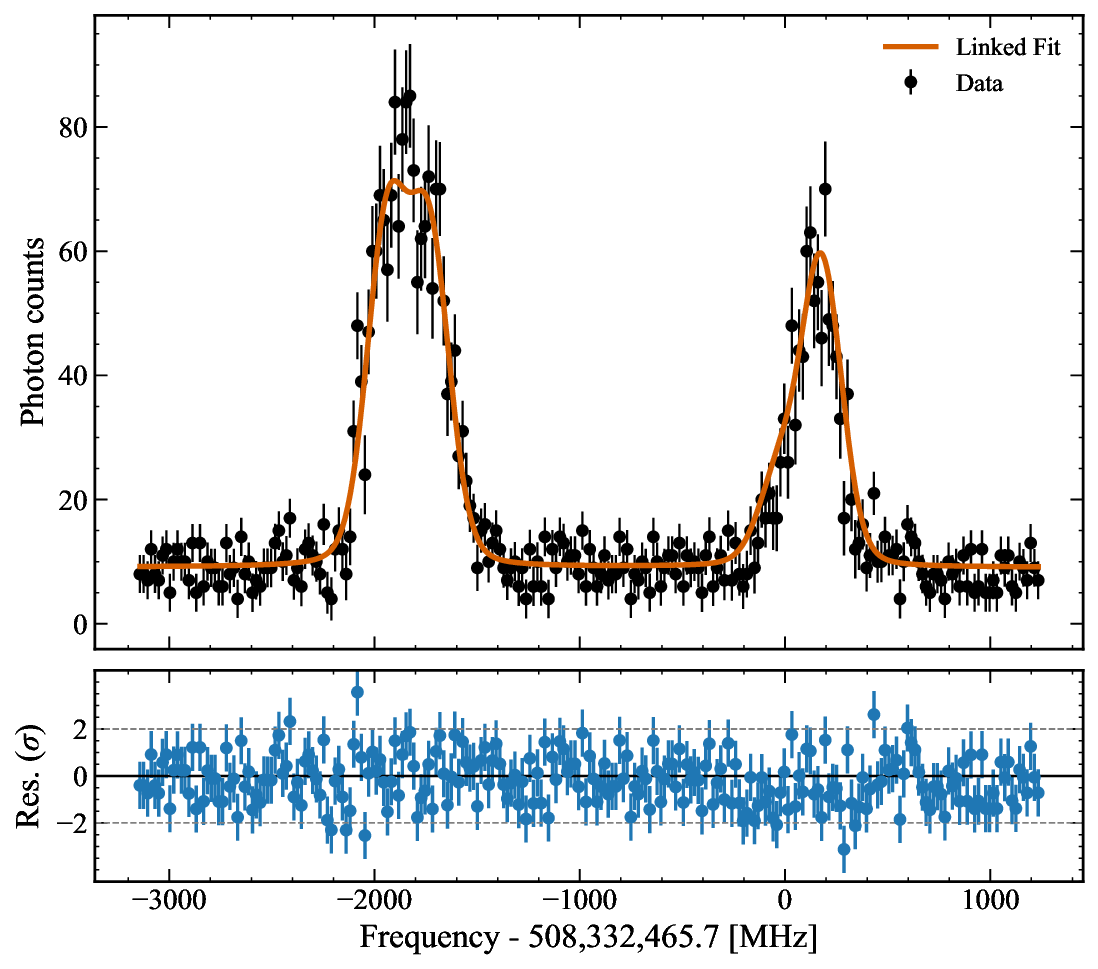}
        \label{fig:21na_anti}
    }
    \caption{\label{fig:21Na_hfs} 
    Same as Fig.~\ref{fig:23Na_hfs} with $^{21}$Na. The relative peak heights follow that of $^{23}$Na in the fitting (see \ref{systematics} for details). The panels show two scans selected from the full dataset of 35 scans.
    }

\end{figure*}

Before performing the measurement with a radioactive beam, the hyperfine structure of the stable isotope $^{23}\text{Na}$ ($I=3/2$) was measured as a reference. A bunched ion beam was delivered from the RFQ-CB with an intensity varying between $1.0 \times 10^9$ and $1.4 \times 10^{10}$ particles per second (pps) and a repetition rate of $10~\text{Hz}$ or $50~\text{Hz}$. The collected dataset consists of 10 scans in anti-collinear and collinear geometries.

The spectra were analyzed using the \texttt{SATLAS2} Python package~\cite{gins2024}. To ensure a consistent treatment of the full dataset, a simultaneous linked fit was performed, in which the magnetic dipole hyperfine constants $A(3s\,^2S_{1/2})$ and $A(3p\,^2P_{1/2})$ were treated as global parameters shared by all scans. Each resonance component was modeled with an asymmetric (skewed) Voigt profile to account for the observed line-shape asymmetry, likely originating from residual beam-energy distortions after RFQ-CB extraction and/or inelastic processes in the charge-exchange cell, and thereby to reduce bias in the extracted hyperfine parameters~\cite{klose2013}. The relative intensities of the hyperfine components were treated as free parameters rather than fixed to the theoretical Racah ratios, since the observed peak amplitudes can be influenced by effects such as optical pumping and residual line-shape distortions~\cite{huber1978,touchard1982}. 

Because the hyperfine peaks are not clearly resolved, the Lorentzian contribution to the line shape could not be constrained robustly from the spectra alone and was therefore not treated as a free spectroscopic parameter. Instead, FWHM$_{L}$ was fixed to a representative working value of 37 MHz, informed by typical operating conditions in the interaction region, including a laser power of about $330~\mu\text{W}$ and an estimated beam diameter of about 0.7 mm. The sensitivity to this assumption was evaluated separately and propagated as a systematic uncertainty, as described in Sec.~\ref{systematics}.

Figure~\ref{fig:23Na_hfs} shows a representative hyperfine spectrum of the $^{23}\text{Na}$ D$_1$ transition. The extracted hyperfine constants are $A(3p\,^2P_{1/2}) = 95.9(10)_{\mathrm{stat}}(2)_{\mathrm{syst}}~\mathrm{MHz}$ and $A(3s\,^2S_{1/2}) = 884.9(16)_{\mathrm{stat}}(4)_{\mathrm{syst}}~\mathrm{MHz}$. These values are consistent with the literature values within the combined statistical and systematic uncertainties~\cite{yei1993,arimondo1977}.

\begin{table*}[t]
\caption{\label{tab:results} Magnetic dipole hyperfine structure constants $A(3p\,^2P_{1/2})$ and $A(3s\,^2S_{1/2})$ for $^{23}\text{Na}$ and $^{21}\text{Na}$ obtained in this work, compared with literature values. Uncertainties are quoted as $(\cdot)_{\mathrm{stat}}(\cdot)_{\mathrm{syst}}$ (see Sec.~\ref{systematics}).}
\begin{ruledtabular}
\begin{tabular}{llll}
Isotope & Dataset / Reference & $A(3p\,^2P_{1/2})$ (MHz) & $A(3s\,^2S_{1/2})$ (MHz) \\
\hline
\noalign{\vspace{2pt}}
$^{23}\text{Na}$ & This work (10 scans)
& $95.9(10)_{\mathrm{stat}}(2)_{\mathrm{syst}}$
& $884.9(16)_{\mathrm{stat}}(4)_{\mathrm{syst}}$
\\
& Yei \textit{et al.}~\cite{yei1993}, Arimondo \textit{et al.}~\cite{arimondo1977}
& $94.44(13)$
& $885.813\,064\,4(5)$ \\

\hline
\noalign{\vspace{2pt}}
$^{21}\text{Na}$ & This work (35 scans)
& $103.6(10)_{\mathrm{stat}}(9)_{\mathrm{syst}}$
& $954.9(11)_{\mathrm{stat}}(25)_{\mathrm{syst}}$
 \\
& Touchard \textit{et al.}~\cite{touchard1982}
& $101.3(23)$
& $954.1(23)$ \\
& Huber \textit{et al.}~\cite{huber1978}
& 102.6(18)
& 953.7(20) \\
& Ames \textit{et al.}~\cite{1965-Ames}
& 
& 953.233(11) \\
\end{tabular}
\end{ruledtabular}
\end{table*}

\subsection{Hyperfine structure of $^{21}\text{Na}$}
Following the reference measurement, the hyperfine structure of the radioactive isotope $^{21}\text{Na}$ ($I=3/2$) was investigated. The ion beam was delivered with an intensity ranging from $1.0 \times 10^7$ to $2.75 \times 10^7$ pps at a repetition rate of $10~\text{Hz}$. The dataset comprises 35 scans in total, including 16 scans in anti-collinear geometry and 19 scans in collinear geometry.

The spectra were analyzed within the same fitting framework used for $^{23}\text{Na}$, including the skewed-Voigt description of each hyperfine component and a simultaneous linked fit with global hyperfine constants shared across all scans. As in the $^{23}\mathrm{Na}$ reference analysis, $\mathrm{FWHM}_L$ was fixed in each fit rather than treated as a free parameter, and the associated model dependence was propagated as a systematic uncertainty, as discussed in Sec.~\ref{systematics}.

For the $^{21}\text{Na}$ analysis, the relative intensity pattern of the hyperfine components was fixed using the $^{23}\text{Na}$ reference data measured in the same transition and analyzed within the same fitting framework. Since both isotopes were studied under the same spectroscopic conditions, this provided an empirical intensity pattern that can reflect deviations from the ideal Racah ratios. The possible model dependence associated with this choice was evaluated separately and included in the systematic uncertainty budget, as described in Sec.~\ref{systematics}.

Figure~\ref{fig:21Na_hfs} shows a representative hyperfine spectrum of the $^{21}\text{Na}$ D$_1$ transition. The extracted hyperfine constants are $A(3p\,^2P_{1/2}) = 103.6(10)_{\mathrm{stat}}(9)_{\mathrm{syst}}~\mathrm{MHz}$ and $A(3s\,^2S_{1/2}) = 954.9(11)_{\mathrm{stat}}(25)_{\mathrm{syst}}~\mathrm{MHz}$. The measured values are in good agreement with the previous studies using laser spectroscopy~\cite{huber1978,touchard1982}.

\subsection{Systematic uncertainties}
\label{systematics}

The main sources of systematic uncertainty in the extracted hyperfine constants of $^{21,23}\text{Na}$ were investigated. In the present analysis, the dominant uncertainties arise from model dependence in the fitting procedure. For both isotopes, this is associated primarily with the adopted line-shape description, while for $^{21}\text{Na}$ an additional contribution arises from the treatment of the relative intensities. Other possible effects, including the acceleration-voltage read-back, residual laser-frequency fluctuations, and geometric misalignment, were also examined and found to be negligible at the present level of precision.

\noindent\textbf{(i) Line-shape model.}
The extracted hyperfine constants may depend on the adopted line-shape model, including the balance between Gaussian and Lorentzian broadening and the treatment of the observed asymmetry. In the nominal analysis, the Gaussian width was left free, whereas $\mathrm{FWHM}_L$ was fixed because it could not be constrained robustly from the unresolved spectra alone. To estimate the associated model dependence, the full dataset was refitted while varying FWHM$_L$ over $26.4$--$53.1~\mathrm{MHz}$, corresponding to the estimated power-broadened Lorentzian widths under the laser-power and beam-diameter conditions described in Sec. II. The maximum absolute deviation of the extracted hyperfine constants from the nominal result was assigned as the systematic uncertainty associated with the Lorentzian contribution to the line shape. For $^{23}\text{Na}$, this procedure yields systematic uncertainties of $0.15~\mathrm{MHz}$ for $A(3p\,^2P_{1/2})$ and $0.42~\mathrm{MHz}$ for $A(3s\,^2S_{1/2})$. For $^{21}\text{Na}$, the corresponding contributions are $0.75~\mathrm{MHz}$ and $0.76~\mathrm{MHz}$, respectively.

\noindent\textbf{(ii) Relative-intensity assumptions in $^{21}\text{Na}$.}
For $^{21}\text{Na}$, the sensitivity of the extracted hyperfine constants to the treatment of the relative intensities was examined using three fitting strategies: intensities fixed to the ideal Racah ratios, intensities constrained by the $^{23}\text{Na}$ reference data, and fully free intensities. The $^{23}\text{Na}$-constrained solution was adopted as the nominal choice. The full spread of the results obtained with these three strategies was used to quantify the model dependence, and its half-width was assigned as the corresponding systematic uncertainty. This gives $0.52~\mathrm{MHz}$ for $A(3p\,^2P_{1/2})$ and $2.34~\mathrm{MHz}$ for $A(3s\,^2S_{1/2})$.

\noindent\textbf{(iii) Combined systematic uncertainty in $^{21}\text{Na}$.}
For $^{21}\text{Na}$, the line-shape- and intensity-related contributions were added in quadrature, resulting in total systematic uncertainties of $0.91~\mathrm{MHz}$ for $A(3p\,^2P_{1/2})$ and $2.46~\mathrm{MHz}$ for $A(3s\,^2S_{1/2})$.

\noindent\textbf{(iv) Minor contributions.}
Possible contributions from the acceleration-voltage read-back, laser-frequency fluctuations, and geometric uncertainties were not included in the final budget. The effect of beam-energy uncertainty is strongly suppressed in the combined collinear and anti-collinear analysis, residual laser-frequency fluctuations are effectively included in the scan-to-scan statistical scatter of the linked fit, and geometric effects are expected to be negligible under the present experimental conditions.

\subsection{Theoretical results}

We calculate the $A(3p\,^2P_{1/2})$ and $A(3s\,^2S_{1/2})$ constants of $^{21}$Na and $^{23}$Na by employing relativistic coupled-cluster theory (RCC). Using the RCC theory {\it ansatz}, we express wave functions of the $3p\,^2P_{1/2}$ and $3s\,^2S_{1/2}$ states as~\cite{Lindgren,Bijaya1} 
\begin{eqnarray}
|\Psi_v \rangle = e^T \{ 1+S_v \} | \Phi_v \rangle ,
\end{eqnarray}
where the reference state is defined as $| \Phi_v \rangle=a_v^{\dagger} | \Phi_0 \rangle$ with the Dirac-Hartree-Fock (DHF) wave function $| \Phi_0 \rangle$ of the closed-shell $[2p^6]$ and the valence orbital $v$. Furthermore, $T$ is the excitation operator for $| \Phi_0 \rangle$ and $S_v$ is the excitation operator for $| \Phi_v \rangle$ due to electron correlation effects.

We accounted for contributions from the Dirac-Coulomb (DC) Hamiltonian first, then added corrections from the Breit, vacuum polarization (VP), and self-energy (SE) effects subsequently in the evaluation of the $A(3p\,^2P_{1/2})$ and $A(3s\,^2S_{1/2})$ constants~\cite{Bijaya2}. We also analyzed corrections to these constants due to the Bohr-Weisskopf (BW) effect~\cite{Bijaya3}.
To demonstrate the need to account for electron correlation effects in order to obtain accurate results, we employed the RCC method at the singles and doubles approximation (RCCSD method), then employed the RCC method at the singles, doubles, and triples approximation (RCCSDT method). We estimate small corrections due to the Breit, VP, SE, and BW effects using the RCCSD method to carry out all these analyses with the available computational resources within the stipulated time frame. Computational details for hyperfine structure constants using our RCC method can be found in Refs.~\cite{Bijaya4,Bijaya5}.

We considered DC contributions through the RCCSD and RCCSDT methods considering atomic orbitals up to $g-$ angular momentum symmetry to demonstrate the importance of triple excitations. Then, we carried out calculations using the DC Hamiltonian including orbitals up to $j-$symmetry in the RCCSD method to account for correlations due to electrons from orbitals with larger angular momentum. The differences in the RCCSD values from both sets of bases are quoted as the ``base contribution". All these estimated contributions along with the final values for $A(3p\,^2P_{1/2})$ and $A(3s\,^2S_{1/2})$ of both $^{21}$Na and $^{23}$Na isotopes are listed in Table \ref{tab1}. The uncertainties quoted with the final values are estimated from the neglected contributions from triples and higher-level excitations. We have used the magnetic dipole moments of both isotopes $\mu(^{21}$Na$)=2.383630\, \mu_N$ and $\mu(^{23}$Na$)=2.2176556\, \mu_N$~\cite{Stone} to estimate the $A$ values. 

\begin{table}[t]
\caption{\label{tab1} Calculated values of the hyperfine structure constants $A$ (in MHz) of the $3s\,^2S_{1/2}$ and $3p\,^2P_{1/2}$ states of the $^{21}$Na and $^{23}$Na isotopes at different levels of approximations. Uncertainties to the final values are quoted within the parentheses.}
\begin{ruledtabular}
\begin{tabular}{lrrrr}
Method &  \multicolumn{2}{c}{$^{21}\text{Na}$} &  \multicolumn{2}{c}{$^{23}\text{Na}$} \\
\cline{2-3} \cline{4-5} \\
     &  $A(3s\,^2S_{1/2})$ & $A(3p\,^2P_{1/2})$  &  $A(3s\,^2S_{1/2})$  & $A(3p\,^2P_{1/2})$ \\
\hline \\
& \\
DHF  & 671.44 &  68.26 &  623.97 & 63.43 \\
RCCSD & 949.45 & 100.43 & 882.19 & 93.31 \\
RCCSDT & 958.57 & 101.83 & 890.83 & 94.63 \\
Basis & $-0.36$ & $-0.03$ & $-0.34$ & $-0.03$ \\
BW & $-0.34$  & $\sim 0.00$ & $-0.31$ & $\sim 0.00$\\
Breit & $-0.59$ & $-0.08$ & $-0.55$ & $-0.08$ \\
VP &  $-0.84$ & $-0.11$ & $-0.78$ &  $-0.11$ \\
SE &  $-4.15$ &  0.03 & $-3.86$ & 0.03 \\
\hline \\
Final [MHz] & 952.29(70) & 101.64(27) & 884.99(65) & 94.44(25) \\
\end{tabular}
\end{ruledtabular}
\end{table}

\begin{figure}[t!]
    \centering
    \includegraphics[width=0.99\linewidth]
    {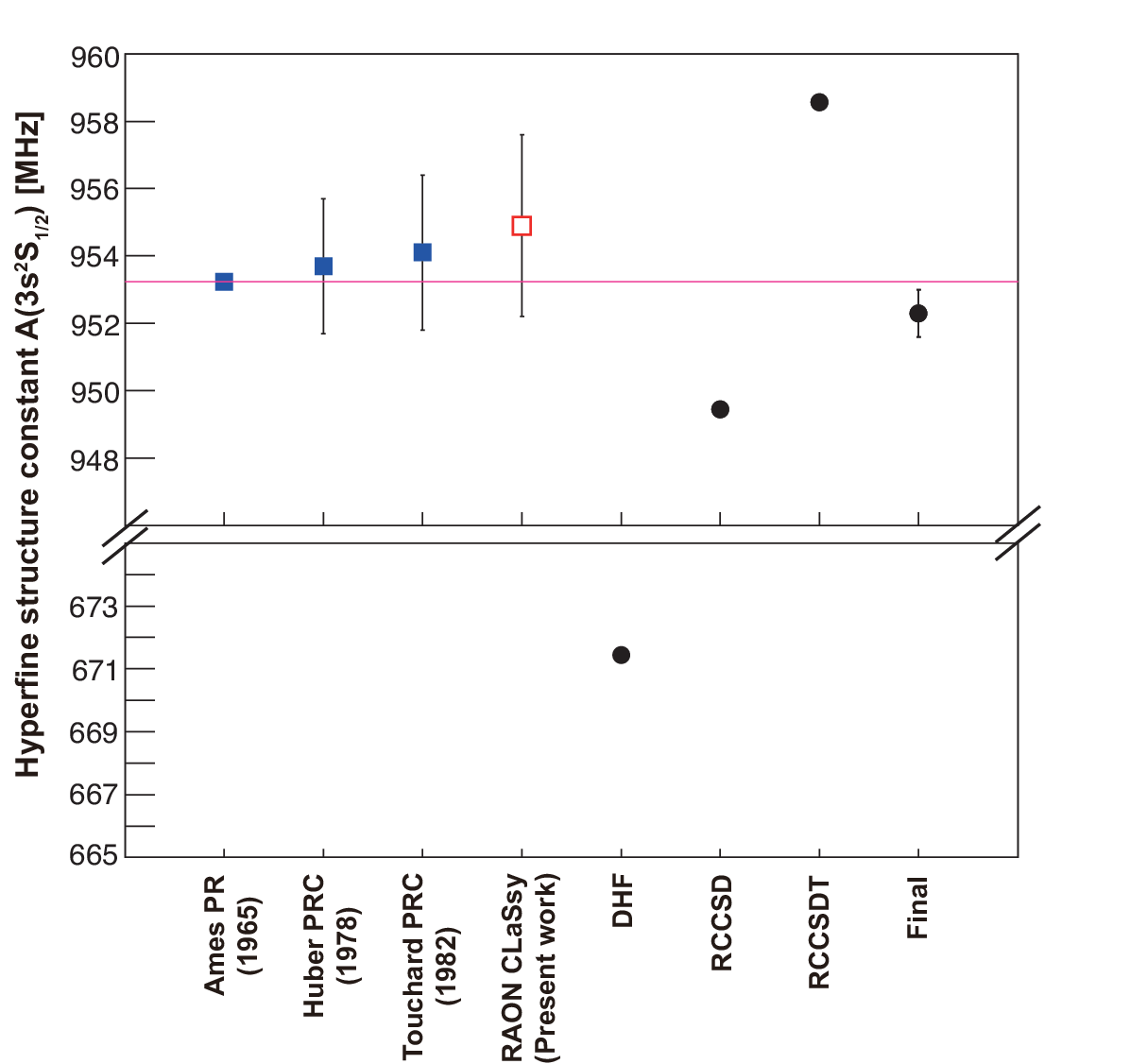}
    \caption{Hyperfine structure constant $A(3s^2S_{1/2})$ of $^{21}$Na from the previous measurements (blue squares)~\cite{1965-Ames,huber1978,touchard1982}, and the measurement at RAON (red square). The non-relativistic mean-field calculation (DHF), the relativistic coupled-cluster calculations (RCCSD and RCCSDT), and the final result considering QED effect are given for reference. The magenta line indicates the $\pm1\sigma$ range of the weighted mean of the literature values and present study, which is dominated by Ref.~\cite{1965-Ames}. 
    }
    \label{fig:A_2S1_2}
\end{figure}

\begin{figure}[t!]
    \centering
    \includegraphics[width=0.99\linewidth]
    {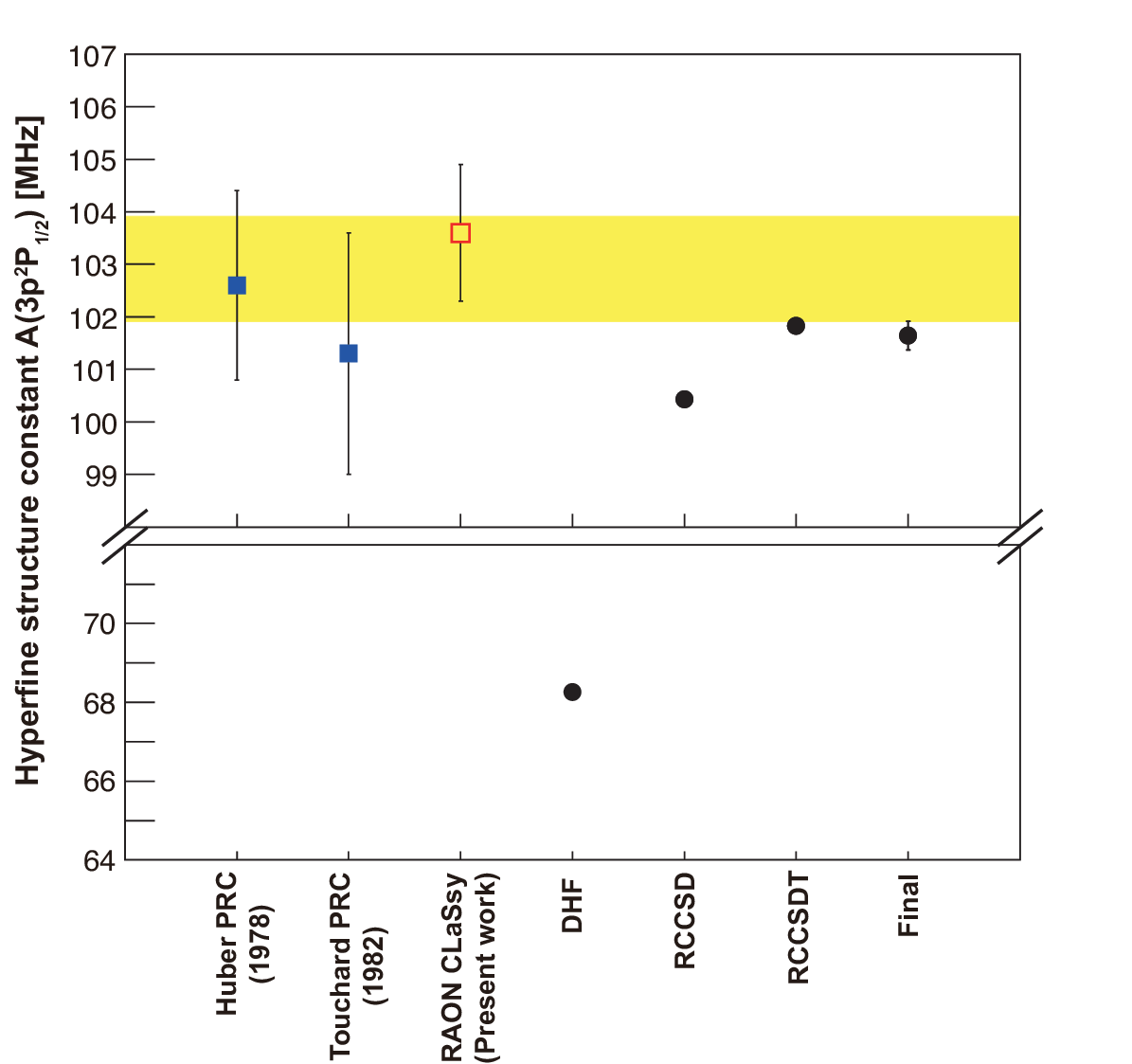}
    \caption{Hyperfine structure constant $A(3p^2P_{1/2})$ of $^{21}$Na from the previous measurements (blue squares)~\cite{huber1978,touchard1982}, and the measurement at RAON (red square). The non-relativistic mean-field calculation (DHF), the relativistic coupled-cluster calculations (RCCSD and RCCSDT), and the final result considering QED effect are given for reference. The yellow band indicates a $\pm1\sigma$ range of the weighted mean of the literature values and present study. 
    }
    \label{fig:A_2P1_2}
\end{figure}

\subsection{Comparison between experimental and theoretical data}

Figure \ref{fig:A_2S1_2} presents both the experimental and theoretical values of the hyperfine structure constant $A(3s^2S_{1/2})$ for $^{21}$Na. As discussed earlier, the QED effects, specifically the VP and SE corrections, reduce the $A(3s^2S_{1/2})$ value obtained from RCCSDT calculations by approximately 5 MHz,
helping to bring the theoretical results into closer agreement with the experimental values. This indicates that the wave function of the ground state has a strong overlap with the nuclear region. Additional corrections, including basis set limitations, Breit interaction, and the BW effect, are also found to be non-negligible; contributing a combined correction of approximately $-1.29$ MHz.

Figure \ref{fig:A_2P1_2} presents both the experimental and theoretical hyperfine structure constant $A(3p^2P_{1/2})$ of $^{21}$Na. The weighted mean of the compiled values is 102.9(10) MHz, representing an improvement over previous measurements reported in Refs.~\cite{huber1978,touchard1982}, which yielded 102.0(14) MHz. This comparison indicates that the compiled value, including our new measurement, clearly distinguishes between the RCCSD and RCCSDT results -- suggesting pivotal roles played by electron correlation effects arising through triple excitations. It can also be noted that QED effects are negligibly small in this case. The very good agreement between the theoretical and experimental results for the ground and first excited states suggests that electron–electron correlation and QED effects are adequately accounted for in our calculations.

\section{Summary and Perspectives}
\label{sec-summary}
To summarize, the hyperfine structure constants $A(3s^2S_{1/2})$ and $A(3p^2P_{1/2})$ of $^{21}$Na were measured at RAON. We employed the CLaSsy setup and an RFQ-CB to obtain the spectra with a high signal-to-noise ratio. 

The measured hyperfine structure constants of the stable Na isotope $^{23}$Na show consistency with the literature values. For $^{21}$Na, the newly measured hyperfine structure constant $A(3p^2P_{1/2})$ exhibit a reduced uncertainty compared to the previous studies. The experimental results demonstrate that the CLaSsy setup is capable to study the hyperfine structure with a radioactive beam. An improvement of the resolution and an increase of detection efficiency will enable the studies of hyperfine structure for more exotic nuclei.

From the comparison with the RCC calculation, we found that incorporating triple excitations—despite their computational complexity—significantly enhances the accuracy of theoretical results, achieving agreement with experimental values. Future investigations aiming for sub-MHz precision will enable more rigorous validation and benchmarking of atomic calculations, opening new avenues for precision atomic physics.

\begin{acknowledgments}
We are thankful to the RAON accelerator team for providing a high-quality ion beam. We acknowledge the assistance of J. Y. Moon for the confirmation of the beam purity. This work was supported by the Institute for Basic Science (IBS-R031-D1 and IBS-R031-Y3). Y.S. acknowledges support from the National Research Foundation of Korea (NRF) grant funded by the Korean government (MSIT) (RS-2022-NR070152). This work was supported by the NRF funded by the MSIT (RS-2024-00436392). The work of BKS at the Physical Research Laboratory (PRL) is supported by ANRF with grant no. CRG/2023/002558 and the Department of Space, Government of India. Atomic calculations were performed on the ParamVikram-1000 HPC cluster at PRL, Ahmedabad, Gujarat, India.
\end{acknowledgments}

\bibliography{21Na_hyperfine_PRA}

@PREAMBLE{
 "\providecommand{\noopsort}[1]{}" 
 # "\providecommand{\singleletter}[1]{#1}%" 
}

@article{1965-Ames,
  title = {Spin, Hyperfine Structure, and Nuclear Magnetic Dipole Moment of 23-sec ${\mathrm{Na}}^{21}$},
  author = {Ames, O. and Phillips, E. A. and Glickstein, S. S.},
  journal = {Phys. Rev.},
  volume = {137},
  issue = {5B},
  pages = {B1157--B1163},
  numpages = {0},
  year = {1965},
  month = {Mar},
  publisher = {American Physical Society},
  doi = {10.1103/PhysRev.137.B1157},
  url = {https://link.aps.org/doi/10.1103/PhysRev.137.B1157}
}

@article{yang2023laser,
  title={Laser spectroscopy for the study of exotic nuclei},
  author={Yang, Xiaofei F and Wang, SJ and Wilkins, SG and Ruiz, RF Garcia},
  journal={Prog. Part. Nucl. Phys.},
  volume={129},
  pages={104005},
  year={2023},
  publisher={Elsevier},
  url={https://www.sciencedirect.com/science/article/pii/S0146641022000631}
}

@article{choi2025,
  title={The RAON facility: an overview},
  author={Choi, Sukjin and Chung, Yeon Sei and Hong, In-Seok and Lee, Jinho and Jung, Yoochul and Kim, Youngman and Kwon, Myeun and Shin, Taeksu and Hong, Seung-Woo},
  journal={J. Korean Phys. Soc.},
  volume={87},
  number={5},
  pages={447--454},
  year={2025},
  publisher={Springer},
  url={https://doi.org/10.1007/s40042-025-01435-1}
}

@article{chung2023,
  title={Commissioning preparation of RAON rare isotope accelerator facility},
  author={Chung, Yeon-Sei and Hong, Seung-Woo and Jang, Hyun-Man and Ki, Tae-Kyung and Kim, Hyoung-Jin and Kim, Young Jin and Kwon, Myeun and Lee, Jinho and Tshoo, Kyoungho and Shin, Taeksu},
  journal={J. Korean Phys. Soc.},
  volume={82},
  number={6},
  pages={613--621},
  year={2023},
  publisher={Springer},
  url={https://doi.org/10.1007/s40042-023-00715-y}
}

@article{park2025development,
  title={Development of the collinear laser spectroscopy (CLaSsy) at RAON},
  author={Park, Sung Jong and Jo, Seong Gi and Lim, Chaeyoung and Tshoo, Kyoungho and Ham, Cheolmin and Kim, Dong Geon and Kwak, Donghyun and Pyeun, Seong Jae and Shin, Taeksu and Kim, Jung Bog and others},
  journal={J. Korean Phys. Soc.},
  volume={87},
  number={5},
  pages={649--654},
  year={2025},
  publisher={Springer},
  url={https://doi.org/10.1007/s40042-024-01205-5}
}

@article{lim2026,
  title={Commissioning of the collinear laser spectroscopy (CLaSsy) at RAON},
  author={Lim, Chaeyoung and others},
  journal={J. Instrum.},
  volume={21},
  number={5},
  pages={P05026},
  year={2026},
  publisher={IOP Publishing},
  url={https://iopscience.iop.org/article/10.1088/1748-0221/21/05/P05026}
}

@article{ohayon2022,
  title={Nuclear charge radii of Na isotopes: Interplay of atomic and nuclear theory},
  author={Ohayon, Ben and Ruiz, RF Garcia and Sun, ZH and Hagen, Gaute and Papenbrock, Thomas and Sahoo, Bijaya Kumar},
  journal={Phys. Rev. C},
  volume={105},
  number={3},
  pages={L031305},
  year={2022},
  publisher={APS},
  url = {https://link.aps.org/doi/10.1103/PhysRevC.105.L031305}
}

@article{huber1978,
  title={Spins, magnetic moments, and isotope shifts of Na 2 1- 3 1 by high resolution laser spectroscopy of the atomic D 1 line},
  author={Huber, G and Touchard, F and B{\"u}ttgenbach, S and Thibault, C and Klapisch, R and Duong, HT and Liberman, S and Pinard, J and Vialle, JL and Juncar, Patrick and others},
  journal={Phys. Rev. C},
  volume={18},
  number={5},
  pages={2342},
  year={1978},
  publisher={APS},
  url = {https://link.aps.org/doi/10.1103/PhysRevC.18.2342}
}

@article{touchard1982,
  title={Electric quadrupole moments and isotope shifts of radioactive sodium isotopes},
  author={Touchard, F and Serre, JM and B{\"u}ttgenbach, S and Guimbal, P and Klapisch, R and de Saint Simon, M and Thibault, C and Duong, HT and Juncar, Patrick and Liberman, S and others},
  journal={Phys. Rev. C},
  volume={25},
  number={5},
  pages={2756},
  year={1982},
  publisher={APS},
  url = {https://link.aps.org/doi/10.1103/PhysRevC.25.2756}
}

@article{yeon2025review,
  title={Review of the 70-MeV cyclotron as the ISOL beam driver at RAON: Y. Yeon et al.},
  author={Yeon, Yeong-Heum and Heo, Seongjin and Yim, Hee-Joong and Woo, Hyung-Joo and Yoo, Kyoung-Hun and Jeong, Jae-Won and Hashimoto, Takashi and Hwang, Wonjoo and Park, Dong-Joon and Nam, Shinwoo and others},
  journal={J. Korean Phys. Soc.},
  volume={87},
  number={5},
  pages={596--604},
  year={2025},
  publisher={Springer},
  url={https://doi.org/10.1007/s40042-024-01199-0}
}

@article{hashimoto2024design,
  title={Design and commissioning of the ISOL beamline at the RAON facility},
  author={Hashimoto, T and Yim, HJ and Kim, JH and Park, Y-H and Heo, S and Yoo, KH and Yun, CC and Ishiyama, H and Lee, JH},
  journal={Nucl. Instrum. Methods B},
  volume={556},
  pages={165504},
  year={2024},
  publisher={Elsevier}
}

@article{kaufman1976,
  title={High-resolution laser spectroscopy in fast beams},
  author={Kaufman, SL},
  journal={Optics Comm.},
  volume={17},
  number={3},
  pages={309--312},
  year={1976},
  publisher={Elsevier},
  url = {https://www.sciencedirect.com/science/article/pii/0030401876902674}
}

@article{gins2024,
title = {SATLAS2: An update to the package for analysis of counting data},
journal = {Computer Phys. Comm.},
volume = {297},
pages = {109053},
year = {2024},
issn = {0010-4655},
doi = {https://doi.org/10.1016/j.cpc.2023.109053},
url = {https://www.sciencedirect.com/science/article/pii/S0010465523003983},
author = {W. Gins and B. {van den Borne} and R.P. {de Groote} and G. Neyens}
}

@article{yei1993,
  title={Delayed-detection measurement of atomic Na 3p 2 P 3/2 hyperfine structure using polarization quantum-beat spectroscopy},
  author={Yei, Wo and Sieradzan, A and Havey, MD},
  journal={Phys. Rev. A},
  volume={48},
  number={3},
  pages={1909},
  year={1993},
  publisher={APS},
  url = {https://link.aps.org/doi/10.1103/PhysRevA.48.1909}
}

@article{arimondo1977,
  title={Experimental determinations of the hyperfine structure in the alkali atoms},
  author={Arimondo, Ennio and Inguscio, M and Violino, P},
  journal={Rev. Mod. Phys.},
  volume={49},
  number={1},
  pages={31},
  year={1977},
  publisher={APS},  
  url = {https://link.aps.org/doi/10.1103/RevModPhys.49.31}
}

@BOOK{Lindgren,
   author       = {I. Lindgren and J. Morrison},
   year         = 1982,
   title        = {Atomic Many-body Theory},
   publisher    = {Springer-Verlag}
}

@article{Bijaya1,
  title={Ab initio determination of the lifetime of the $6p~ {^2}P_{3/2}$ state for $^{207}$Pb$^+$ by relativistic many-body theory},
  author={Sahoo, Bijaya K and Majumder, Sonjoy and Chaudhuri, Rajat K and Das, BP and Mukherjee, Debashis},
  journal={J. Phys. B: At., Mol. and Opt. Phy.},
  volume={37},
  number={17},
  pages={3409},
  year={2004},
  publisher={IOP Publishing},
  url = {https://doi.org/10.1088/0953-4075/37/17/002}
}

@article{Bijaya2,
  title={Conforming the measured lifetimes of the $5d~ {^2}D_{3/2,5/2}$ states in Cs with theory},
  author={Sahoo, B. K.},
  journal={Phys. Rev. A},
  volume={93},
  number={2},
  pages={022503},
  year={2016},
  publisher={APS},
  doi = {10.1103/PhysRevA.93.022503},
  url = {https://link.aps.org/doi/10.1103/PhysRevA.93.022503}
}

@article{Bijaya3,
  title={Precise determination of electric quadrupole moments and isotope shift constants of Yb$^+$ in pursuance of probing fundamental physics and nuclear radii},
  author={Sahoo, B. K.},
  journal={Phys. Rev. A},
  volume={111},
  number={6},
  pages={L060801},
  year={2025},
  publisher={APS},
  doi = {10.1103/PhysRevA.111.L060801},
  url = {https://link.aps.org/doi/10.1103/PhysRevA.111.L060801}
}

@article{Sahoo2003,
  title = {Magnetic dipole hyperfine interactions in ${}^{137}{\mathrm{Ba}}^{+}$ and the accuracies of the neutral weak interaction matrix elements},
  author = {Sahoo, Bijaya K. and Gopakumar, Geetha and Chaudhuri, Rajat K. and Das, B. P. and Merlitz, Holger and Mahapatra, Uttam Sinha and Mukherjee, Debashis},
  journal = {Phys. Rev. A},
  volume = {68},
  issue = {4},
  pages = {040501},
  numpages = {4},
  year = {2003},
  month = {Oct},
  publisher = {American Physical Society},
  doi = {10.1103/PhysRevA.68.040501},
  url = {https://link.aps.org/doi/10.1103/PhysRevA.68.040501}
}

@article{Bijaya4,
  title={Application of relativistic coupled-cluster theory to heavy atomic systems with strongly interacting configurations: Hyperfine interactions in $^{207}$Pb$^+$},
  author={Sahoo, Bijaya K and Chaudhuri, Rajat K and Das, B P and Merlitz, Holger and Mukherjee, Debashis},
  journal={Phys. Rev. A},
  volume={72},
  number={3},
  pages={032507},
  year={2005},
  publisher={APS},
  url = {https://link.aps.org/doi/10.1103/PhysRevA.72.032507}
}

@article{Sahoo2018,
  title = {Relativistic coupled-cluster-theory analysis of energies, hyperfine-structure constants, and dipole polarizabilities of ${\mathrm{Cd}}^{+}$},
  author = {Li, Cheng-Bin and Yu, Yan-Mei and Sahoo, B. K.},
  journal = {Phys. Rev. A},
  volume = {97},
  issue = {2},
  pages = {022512},
  numpages = {9},
  year = {2018},
  month = {Feb},
  publisher = {American Physical Society},
  doi = {10.1103/PhysRevA.97.022512},
  url = {https://link.aps.org/doi/10.1103/PhysRevA.97.022512}
}

@article{Voltka,
  title = {Test of Many-Electron QED Effects in the Hyperfine Splitting of Heavy High-$Z$ Ions},
  author = {Volotka, A. V. and Glazov, D. A. and Andreev, O. V. and Shabaev, V. M. and Tupitsyn, I. I. and Plunien, G.},
  journal = {Phys. Rev. Lett.},
  volume = {108},
  issue = {7},
  pages = {073001},
  numpages = {5},
  year = {2012},
  month = {Feb},
  publisher = {American Physical Society},
  doi = {10.1103/PhysRevLett.108.073001},
  url = {https://link.aps.org/doi/10.1103/PhysRevLett.108.073001}
}

@article{Bijaya5,
  title={Appraising nuclear-octupole-moment contributions to the hyperfine structures in Fr 211},
  author={Sahoo, B. K.},
  journal={Phys. Rev. A},
  volume={92},
  number={5},
  pages={052506},
  year={2015},
  publisher={APS},
  doi = {10.1103/PhysRevA.92.052506},
  url = {https://link.aps.org/doi/10.1103/PhysRevA.92.052506}
}

@article{Stone,
  title={Table of nuclear magnetic dipole and electric quadrupole moments},
  author={Stone, N. J.},
  journal={At. Data Nuc. Data Tab.},
  volume={90},
  number={1},
  pages={75--176},
  year={2005},
  publisher={Elsevier},
  doi = {https://doi.org/10.1016/j.adt.2005.04.001},
  url = {https://www.sciencedirect.com/science/article/pii/S0092640X05000239},
}

@article{pescht1977isotope,
  title={Isotope shift and HFS of D 1 lines in Na-22 and 23 measured by saturation spectroscopy},
  author={Pescht, K and Gerhardt, H and Matthias, E},
  journal={Z. Phys. A},
  volume={281},
  number={3},
  pages={199--204},
  year={1977},
  publisher={Springer},
  url={https://doi.org/10.1007/BF01408837}
}

@article{campbell2016laser,
  title={Laser spectroscopy for nuclear structure physics},
  author={Campbell, P and Moore, Iain D and Pearson, Matthew R},
  journal={Prog. Part. Nucl. Phys.},
  volume={86},
  pages={127--180},
  year={2016},
  publisher={Elsevier},
  url={https://www.sciencedirect.com/science/article/pii/S0146641015000915}
}

@misc{nndc2026,
  author  = {NNDC},
  title   = {NuDat 3.0},
  howpublished = {https://www.nndc.bnl.gov/nudat3/},
  year         = {2026},
  note         = {Accessed: 2026-02-13}
}

@article{koszorus2024nuclear,
  title={Nuclear structure studies by collinear laser spectroscopy},
  author={Koszor{\'u}s, {\'A} and De Groote, RP and Cheal, B and Campbell, P and Moore, ID},
  journal={Eur. Phys. J. A},
  volume={60},
  number={1},
  pages={20},
  year={2024},
  publisher={Springer},
  url={https://link.springer.com/article/10.1140/epja/s10050-024-01230-9}
}

@article{klose2013,
  title={Collinear laser spectroscopy on the ground state and an excited state in neutral $^{55}$Mn},
  author={Klose, A. and Minamisono, K. and Mantica, P. F.},
  journal={Phys. Rev. A},
  volume={88},
  pages={042701},
  year={2013},
  publisher={American Physical Society},
  doi={10.1103/PhysRevA.88.042701}
}

@BOOK{Foot,
   author       = {C. J. Foot},
   year         = 2005,
   title        = {Atomic Physics},
   publisher    = {Oxford University Press}
}

@misc{steck2003,
  author  = {D. A. Steck},
  title   = {Sodium D Line Data},
  howpublished = {https://steck.us/alkalidata/sodiumnumbers.1.6.pdf},
  year         = {2003},
  note         = {Accessed: 2026-05-21}
}

\end{document}